\documentclass[aps, pra, twocolumn, superscriptaddress]{revtex4-2}
\usepackage{placeins}
\usepackage{amssymb, amsmath, amsthm}
\usepackage[utf8]{inputenc}
\usepackage[english]{babel}
\usepackage{color}
\usepackage{graphicx}
\usepackage{hyperref}
\usepackage{comment}
\usepackage{mathtools}
\usepackage{color}
\usepackage{soul}
\usepackage{mathtools}
\usepackage{float}
\usepackage{bm}
\newcommand{\be}{\begin{equation}}
\newcommand{\ee}{\end{equation}}
\newcommand{\ba}{\begin{eqnarray}}
\newcommand{\ea}{\end{eqnarray}}

\usepackage{bbm}
\usepackage{tikz}
\usetikzlibrary{backgrounds}

\DeclareMathOperator{\diagonal}{diag}
\DeclareMathOperator{\trace}{Tr}

\newcommand{\trig}[3][]{#2^{#1}\left(#3\right)}
\newcommand{\diff}{\mathrm{d}}

\newcommand{\qeye}[1][]{\mathbbm{1}_{#1}}
\newcommand{\adj}[1]{#1^\dagger}

\renewcommand{\L}{\mathrm{L}}
\newcommand{\R}{\mathrm{R}}
\newcommand{\LR}{{\L,\R}}

\newcommand{\B}{\mathrm{B}}

\newcommand{\sys}{\mathrm{S}}
\newcommand{\eff}{\mathrm{eff}}
\newcommand{\traj}{\mathrm{traj}}
\newcommand{\detec}{\mathrm{d}}

\newcommand{\paulix}{\sigma^{\mathrm{x}}}
\newcommand{\pauliy}{\sigma^{\mathrm{y}}}
\newcommand{\pauliz}{\sigma^{\mathrm{z}}}

\newcommand{\ket}[1]{\left|#1\right\rangle}
\newcommand{\bra}[1]{\left\langle#1\right|}

\newcommand{\diag}[1]{\diagonal\left(#1\right)}
\newcommand{\Tr}[2][]{\trace_{#1}\left[ #2 \right]}

\newcommand{\ev}[1]{\langle#1\rangle}

\newcommand{\tl}{\theta_\L}
\newcommand{\tr}{\theta_\R}
\newcommand{\tlr}{\theta_{\L, \R}}

\definecolor{crimson}{HTML}{DC143C}
\definecolor{coral}{HTML}{FF7F50}

\newcommand{\mytitle}{Theory of the correlated quantum Zeno effect in a monitored qubit dimer}

\graphicspath{{./figures/}}

\begin{document}
  
\title{\mytitle}      

\author{Severino Zeni}
\email{severino.zeni@exact-lab.it}
\affiliation{Pitaevskii BEC Center, CNR-INO and Dipartimento di Fisica, Universit\`a di Trento, I-38123 Trento, Italy}
\affiliation{eXact lab Srl - Janas Project, Via Francesco Crispi 56, 34126 Trieste, Italy}

\author{Gobinda Chakraborty}
\affiliation{Department of Physics, Lancaster University, Lancaster, LA1 4YB, United Kingdom}

\author{Alessandro Romito}
\affiliation{Department of Physics, Lancaster University, Lancaster, LA1 4YB, United Kingdom}

\author{Alberto Biella}
\email{alberto.biella@cnr.it}\affiliation{Pitaevskii BEC Center, CNR-INO and Dipartimento di Fisica, Universit\`a di Trento, I-38123 Trento, Italy}
\affiliation{INFN-TIFPA, Trento Institute for Fundamental Physics and Applications, I-38123 Trento, Italy}

\begin{abstract}
We theoretically investigate the stochastic dynamics of two qubits subject to one- and two-site correlated continuous weak measurements. 
When measurements dominate over the local unitary evolution, the system's dynamics is constrained and part of the physical Hilbert space becomes inaccessible: a typical signature of the Quantum Zeno (QZ) effect. 
In this work, we show how the competition between these two measurement processes give rise to two distinct QZ regimes, we dubbed {\it standard} and {\it correlated}, characterised by a different topology of the allowed region of the physical Hilbert space being a simply and non-simply connected domain, respectively.
We develop a theory based on a stochastic Gutzwiller ansatz for the wavefunction that is able to capture the structure of the phase diagram. Finally we show how the two QZ regimes are intimately connected to the topology of the flow of the underlying non-Hermitian Hamiltonian governing the no-click evolution.
\end{abstract}

\maketitle

\section{Introduction}
A watched pot never boils over, or does it?
According to quantum mechanics, it depends on the rate at which we watch the pot.
In a generic quantum system, the unitary evolution, which tends to delocalize the wavefunction over the entire Hilbert space, competes with quantum measurements, which, on the contrary, hamper the system's dynamics~\cite{petruccioneBook}.
When measurements dominate over the Hamiltonian dynamics, the wavefunction gets localized in a specific region of the Hilbert space, giving rise to the celebrated QZ effect~\cite{misra1977thezeno,peres80zeno,Itano_PRA1990,facchi2001from,facchi2002quantum,Kwiat1999,Wolters2013,signoles2014confined,Facchi_2008,FACCHI2001147,PhysRevA.69.032314,greenfield2025}.

The accurate monitoring of quantum fluctuations of the measuring apparatus leads to the possibility of tracking individual quantum trajectories, i.e. to resolve single instances of the stochastic evolution of the system's wavefunction~\cite{Guerlin2007,Sayrin2011,Murch2013_bis,Murch2013,Weber2014}.
By collecting many trajectories one can construct the statistical mixture of states explored by the system, i.e. the system density matrix.
While the same mixture can be constructed in many different ways, a fact known as {\it ensemble ambiguity}~\cite{nielsenBookQCQI,minganti2024openquantumsystems}, the knowledge of its unravelling in terms of trajectories gives much more informations about the underlying monitored dynamics, including space-time correlations and the full counting statistics of the observables~\cite{GARRAHAN2018130,Landi2024,Fazio2025}.

At a given time the statistics of the trajectories can be condensed in the probability density function (PDF) that provides the probability to find the quantum state in a specific {\it position} of the Hilbert space.
For a single qubit, the structure of the PDF provides a direct proxy for the emergence of the QZ effect featuring forbidden regions where at long times the PDF vanishes~\cite{snizhko2020quantumzeno}.
When more than one body is considered, interactions play a crucial role in determining the onset of the QZ effect~\cite{froml2019fluctuation,froml2020ultracold,Biella2021,Rossini2021,Secli2022,Rosso_SUN_2022,Rosso_fermions_strong_2023,leung2024,Wauters2025}.
Furthermore, on a lattice, correlated multi-site measurements (i.e. the detection of the collective state of the sites involved) are  possible and compete with one-site monitoring~\cite{Ippoliti2021, Piccitto2023}.
In the latter case, the structure of the PDF is generally unknown, and the impact of different detection schemes and measurements remains unexplored.

\begin{figure}[b]
    \begin{minipage}[t]{0.385\linewidth}
        \centering
        (a)
        
        {\footnotesize
        \begin{tikzpicture}[scale=.365]
            \draw[very thick] (0, 2) -- (2, 2) node[right] {$|0_\L\rangle$};
            \draw[very thick] (0, 0) -- (2, 0) node[right] {$|1_\L\rangle$};
            \draw[very thick] (5, 2) -- (7, 2) node[right] {$|0_\R\rangle$};
            \draw[very thick] (5, 0) -- (7, 0) node[right] {$|1_\R\rangle$};
            
            \draw[->] (1.7071, .2929) arc[start angle=-45, end angle=45, radius=1];
            \draw[->] (.2929, 1.7071) arc[start angle=135, end angle=225, radius=1];
            \node at (1, 1) {$\omega_\sys$};
            \draw[->] (6.7071, .2929) arc[start angle=-45, end angle=45, radius=1];
            \draw[->] (5.2929, 1.7071) arc[start angle=135, end angle=225, radius=1];
            \node at (6, 1) {$\omega_\sys$};

            \draw[color=crimson, thick, dashed] (.5, -.5) -- (.5, -4.5) node[black, midway, right] {$\lambda_1$};
            \draw[color=crimson, thick, dashed] (6.5, -.5) -- (6.5, -4.5) node[black, midway, right] {$\lambda_1$};
            \draw[color=coral, thick, dashed] (1.5, -.5) -- (3.5, -4.5);
            \draw[color=coral, thick, dashed] (5.5, -.5) -- (3.5, -4.5);
            \draw[color=coral, thick, dashed] (1.5, -.5) to[out=-45, in=225] (5.5, -.5);
            \node at (3.5, -2.25) {$\lambda_2$};

            \draw[thin] (-.2027, -5.2929) arc[start angle=135, end angle=45, radius=1];
            \filldraw (0, -5.1340) circle (.075);
            \filldraw (1, -5.1340) circle (.075);
            \filldraw (.5, -6) circle (.075); 
            \draw[->, thin] (.5, -6) -- +(120:.8);
            \draw[thin] (5.7929, -5.2929) arc[start angle=135, end angle=45, radius=1];
            \filldraw (6, -5.1340) circle (.075);
            \filldraw (7, -5.1340) circle (.075);
            \filldraw (6.5, -6) circle (.075); 
            \draw[->, thin] (6.5, -6) -- +(120:.8);
            \draw[thin] (2.7929, -5.2929) arc[start angle=135, end angle=45, radius=1];
            \filldraw (3, -5.1340) circle (.075);
            \filldraw (4, -5.1340) circle (.075);
            \filldraw (3.5, -6) circle (.075); 
            \draw[->, thin] (3.5, -6) -- +(120:.8);

        \end{tikzpicture}
        }
    \end{minipage}
    \hfill\vline\hfill
    \begin{minipage}[t]{0.6\linewidth}
        \centering
        (b)
        
        \includegraphics[scale=1]{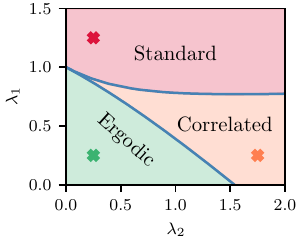}
    \end{minipage}
    \caption{(a) Sketch of the system. Two identical qubits ($\L$ and $\R$) undergo coherent oscillations with frequency $\omega_\sys$ and are subject to one- and two-site continuous weak measurements with effective strength $\lambda_1$ and $\lambda_2$, respectively. 
    (b) Phase diagram of the monitored qubit dimer within the Gutzwiller approximation.
    The crosses represent the selection of parameters that were simulated and are represented in Fig. \ref{fig:gutz_combo} and Fig. \ref{fig:dimer-full-vs-gutz}.}
    \label{fig:sketch+phase-diagram}
\end{figure}

To address these fundamental questions, this study explores a minimal model composed of two qubits subject to one- and two-qubit measurements and undergoing local unitary dynamics. 
We show that the competition between the two kinds of  measurements and Hamiltonian evolution leads to the emergence of two distinct QZ regimes characterized by a different topology of the PDF.
Our main result is that simultaneous two-qubit measurements stabilize a correlated QZ regime where  the topology of the PDF is a non-simply connected domain surrounding a non-accessible region.
On a physical ground, this result establishes that each qubit can be found in any possible state, i.e. it fully spreads over its local Hilbert space, but some configurations of the two qubits are forbidden.
Extending the famous proverb we claim that {\it two pots watched simultaneously never boil together, however they do it separately}.

When one-qubit measurements dominate over the correlated ones, the dynamics of the two qubits effectively decouples, and the system approaches a standard QZ regime where only a simply connected domain of the physical Hilbert space is explored during the dynamics. 
We determine the boundaries of the different phases in the parameter space, developing a theory based on the stochastic Gutzwiller ansatz for the dimer wavefunction~\cite{Jin2016,Huybrechts2020,Wouter2023,ares2024restrictedmontecarlowave}. 
The phase diagram of the model is shown in Fig.\ref{fig:sketch+phase-diagram}.
We also show that the different topology of the PDF is in one-to-one correspondence with the topology of the flow~\cite{Villa2024} of the postselected non-Hermitian no-click evolution.

\section{Model and measurement protocol}
We consider two identical two-level systems each performing coherent oscillations between the states $\ket{0_{\L, \R}}$ and $\ket{1_{\L, \R}}$ ($\L$ and $\R$ refer to the left and right spin, respectively) due to the Hamiltonian $H_\sys= \omega_\sys \left(\paulix_{\L} + \paulix_{\R}\right)$, where $\omega_\sys > 0$ is the oscillation frequency (hereafter we set $\hbar=1$).
The two qubits also be monitored by a sequence of variable strength measurements at an interval $dt \ll 1 / \omega_\sys$.

By coupling the system with ancillary qubits appropriately, it is possible to construct a protocol where both the on-site populations and the two-qubit density correlations are measured. 
A sketch of the setup under consideration is shown in Fig.~\ref{fig:sketch+phase-diagram} (see also Appendix \ref{app:phys_mod} for details of the physical model). 
More formally, each measurement event is characterised by four possible readouts $r = 0, 1, 2, 3$, which give rise to the following back-actions on the system
\begin{equation}
    \begin{split}
        &M_1 = \sqrt{p_1} n_\L, \quad M_2 = \sqrt{p_1} n_\R,\\
        &M_3 = \sqrt{p_2} n_\L n_\R, \\
        &M_0 = \sqrt{\qeye - M_1^2 - M_2^2 - M_3^2}
    \end{split}
    \label{eq:dimer-povm}
\end{equation}
where $n_{\L,\R}=(\qeye-\pauliz_{\L,\R})/2=\ket{1_{\L, \R}}\bra{1_{\L, \R}}$, and $p_1, p_2 \in [0, 1]$ control the measurement strength.
Readouts \mbox{$r=1,2$} account for on-site measurement of the $\L$ and $\R$ qubit population, respectively.
The $r=3$ readout accounts for the correlated measurement of the two qubits.
Finally, the event where none of the detectors click, $r=0$, often referred in the literature as {\itshape no-click} event.
We are interested in the regime where $p_{1, 2} = \gamma_{1, 2} dt$ with $dt \rightarrow 0$ and $\gamma_{1, 2} \geq 0$.
According to the Born rule, the probability of obtaining each of the readouts on a state $\ket{\psi}$ is given by $p_r = \bra{\psi} \adj{M_r} M_r \ket{\psi}$ for $r = 0, 1, 2, 3$ \cite{nielsenBookQCQI}.

In each time step $dt$ the unitary dynamics and measurement combine into the evolution 
\begin{equation}
    \ket{\psi(t + dt)}
    = \frac{M_r U}{\sqrt\mathcal{N}} \ket{\psi(t)}
    \label{eq:dimer-stochastic-evolution}
\end{equation}
where $U = e^{-i H_\sys dt} \approx \qeye - i dt H_{\rm S}$ and $\mathcal{N}$ ensures the normalization of the post-measurement state.

Note that in the monitoring protocol we are considering, the two qubits do not interact because of the unitary evolution, but they are coupled via the readouts $r = 0$ and $r = 3$ of the measurement procedure.
When $r=0$, the no-click dynamics can be written in terms of a non-Hermitian Hamiltonian.
Indeed $M_0 U \approx \qeye - i dt H_{\rm eff}$ where
\begin{equation}
\label{nonHermHam}
    H_{\rm eff} = H_{\rm S} -i \left[\frac{\gamma_1}{2} (n_\L+n_\R) + \frac{\gamma_2}{2}n_\L n_\R \right].
\end{equation}
The last term in Eq. \eqref{nonHermHam} explicitly couples the two qubits and generates entanglement among them.
When $r = 3$, we know that after the measurement the two-qubit state will coincide with the pointer state $\ket{1_\L}\ket{1_\R}$, thus completely correlating (classically) the states of the two qubits.

\section{The stochastic Gutzwiller ansatz}
In order to get further insights about the monitored dynamics, we propose a Gutzwiller ansatz that for the dimer under investigation takes the form
\begin{equation} \label{gw_ansatz}
    \ket{\psi(t)}
    = \ket{\psi_\L(t)} \ket{\psi_\R(t)}.
\end{equation}
The ansatz \eqref{gw_ansatz} allows to capture the build up of classical correlations between the left and right qubit induced by the monitoring dynamics and will neglect the quantum ones due to the fact that the state is always factorisable.
Indeed, the factorized ansatz \eqref{gw_ansatz} cannot capture the entanglement between the two qubits. However, the state of the left/right qubit is affected by the stochastic dynamics of the right/left one, coupling the evolution of the local states.

This approximation is well justified in our problem since clicks ($r = 1, 2, 3$, see Eq. \eqref{eq:dimer-povm}) collapse the system to factorised states.
It is therefore reasonable to assume that the system will remain in the vicinity of a factorised state at all times.
Only the no-click event ($r=0$) couples the two qubits, generating both classical and quantum correlation. 
The extent to which this approximation is justified quantitatively will be evaluated by comparing the approximate results to simulations of the full system dynamics for the quantities of interests.

Very conveniently, in the Gutzwiller approximation the state of the system can be represented on two Bloch spheres, one for each qubit.
In fact, for a suitable set of initial states such that $\bra{\psi(0)}\paulix_{\LR}\ket{\psi(0)}=0$, the dynamics of the system is confined to a section of the two spheres.
Hence, one angle $\theta_{\LR}\in (-\pi, \pi)$ for each qubit is sufficient for describing the state that can be parametrized as $\ket{\psi_{\LR}}
    = \cos \frac{\theta_{\LR}}{2} \ket{0_{\LR}}
      + i \sin \frac{\theta_{\LR}}{2} \ket{1_{\LR}}$.
Thus, the pair of variables $(\tl, \tr)$ fully determine the state of the dimer.

Within the Gutzwiller approximation, the dynamics of the system can then be rewritten as a time evolution of the two variables $\tl$ and $\tr$.
In terms of these Eq. \eqref{eq:dimer-stochastic-evolution} becomes (see Appendix \ref{sec:dime-gutz-derivation} for the details of the derivation):
\begin{equation}
    \begin{pmatrix}
        \tl(t+dt) \\
        \tr(t+dt)
    \end{pmatrix} =
    \begin{cases}
        \begin{pmatrix}
            \tl(t) - \Omega_\L \left( \tl, \tr \right) dt \\
            \tr(t) - \Omega_\R \left( \tl, \tr \right) dt
        \end{pmatrix}
        & \text{if } r = 0 \\[12pt]
        \begin{pmatrix}
            \pi \\
            \tr(t)
        \end{pmatrix}
        & \text{if } r = 1 \\[12pt]
        \begin{pmatrix}
            \tl(t) \\
            \pi
        \end{pmatrix}
        & \text{if } r = 2 \\[12pt]
        \begin{pmatrix}
            \pi \\
            \pi
        \end{pmatrix}
        & \text{if } r = 3 \\
    \end{cases}
    \label{eq:dimer-gutz-stochastic-evolution}
\end{equation}
where $\Omega_\L\left(\tl, \tr\right)
        = 2 \omega_\sys \left[1 + \left(\lambda_1 + \lambda_2 \sin^2\frac{\theta_R}{2} \right) \sin\tl\right]$ and $\Omega_\R\left(\tl, \tr\right) = \Omega_\L\left(\tr, \tl\right)$,
with the effective adimensional measurement strengths \mbox{$\lambda_{1, 2} = \gamma_{1, 2} / (4 \omega_\sys)$}.
Equation \eqref{eq:dimer-gutz-stochastic-evolution}, complemented with the readout probabilities in the Gutzwiller approximation
\begin{equation}
    \begin{split}
        &p_{1, 2} = \gamma_1  \sin^2(\tlr/2) dt,\\
        &p_3 = \gamma_2  \sin^2(\tl/2) \sin^2(\tr/2) dt,\\
        &p_0 = 1 - \left( p_1 + p_2 + p_3 \right),
    \end{split}
    \label{eq:dimer-gutz-readout-probabilities}
\end{equation} 
describes a stochastic evolution of the two variables $\tl$ and $\tr$.
Summarizing: measurements yielding readouts $r = 1 ~(2)$ project the L (R) qubit onto the state $\tl = \pi$ ($\tr = \pi$).
Measurements yielding the readout $r = 3$ take the system to the state $\tl = \tr = \pi$.
While when the readout $r = 0$ (no-click) is obtained the variables $(\tl, \tr)$ evolve infinitesimally with velocity $(\Omega_\L(\tl, \tr), \Omega_\R(\tl, \tr))$.
The Gutzwiller decoupling leads to a non-linear stochastic dynamics for the evolution of the local states.

Finally, we stress that this approach can be easily extended to the $N$-qubit scenario. In such case, this would translate into a set of non-linear stochastic equations for the $N$ angles $(\theta_1,\dots,\theta_N)$
fully parametrizing the state.

\section{Probability density of the quantum state}
The main quantity we are interested in is the PDF $P_t(\ket{\psi})$ that represents the probability for the system of being in the state $\ket{\psi}$ at time $t$ given the ensemble of trajectories generated by the monitored dynamics.
In general, for a system of $N$ spins, this is a function of the $\mathcal{O}\left(2^N\right)$ parameters that are needed to pinpoint a state in the system's Hilbert space.
However, within the Gutzwiller framework, the state of the dimer is fully determined by the variables $(\tl, \tr)$.
Operationally, this function can be written as 
\be
\label{P_def}
P_t(\tl, \tr) = \frac{1}{N_{\rm traj}} \sum_{i=1}^{N_{\rm traj}} 
\prod_{\alpha=\L,\R}\delta\left(\theta_\alpha - \theta_\alpha^{(i)}(t)\right),
\ee
where $\theta_\alpha^{(i)}(t)$ parametrizes the Gutzwiller state along the $i$-th trajectory at time $t$, 
 $N_{\rm traj}$ is the number of trajectories and $\delta(\bullet)$ denotes the Dirac delta distribution.
This PDF directly probes the states' spreading in the portion of the Hilbert space explored by the dynamics.
It is thus a powerful tool to witness the onset the QZ effect, which hampers the state evolution creating forbidden regions in said portion of the Hilbert space.
The master equation governing the dynamics of $P_t(\tl, \tr)$ is derived in Appendix \ref{app:gw_master_eq}.

\begin{figure}[htb]
\centering
    \includegraphics[scale=0.6]{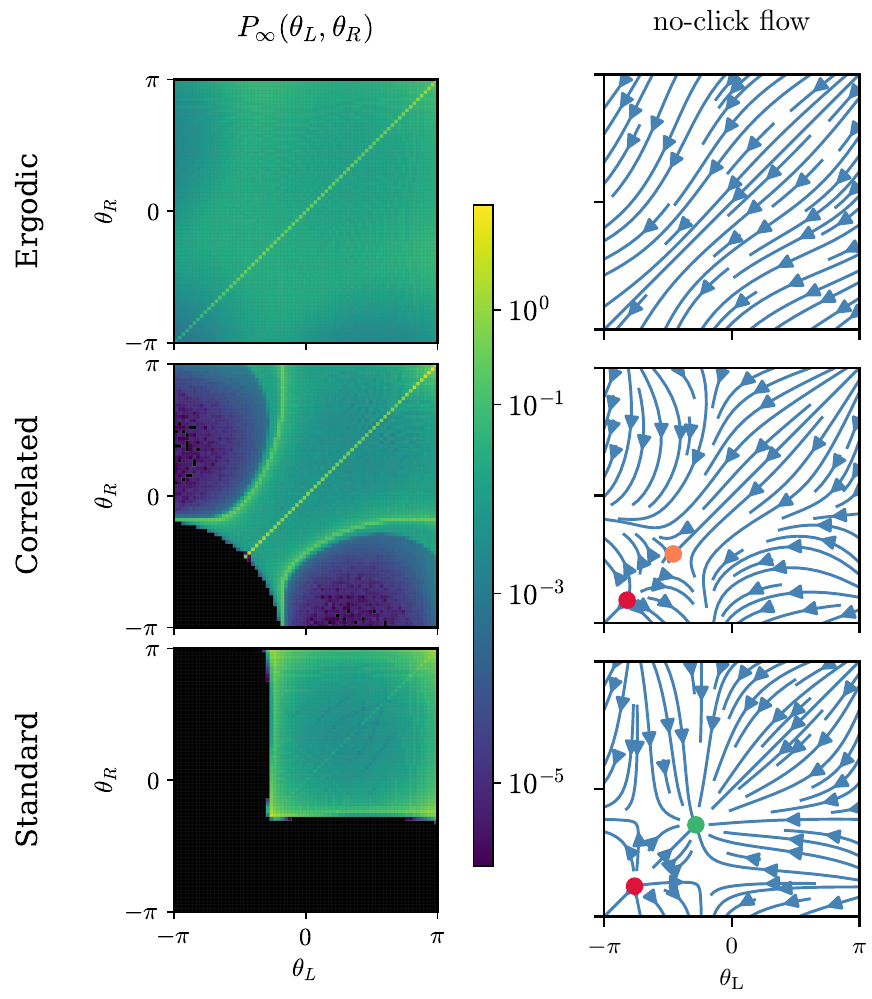}
    \caption{Left panels: Stationary probability distributions $P_\infty(\tl, \tr)$ for measurement strengths $(\lambda_1, \lambda_2) = (0.25, 0.25)$ (top), $(\lambda_1, \lambda_2) = (0.25, 1.75)$ (middle), and$(\lambda_1, \lambda_2) = (1.25, 0.25)$ (bottom). The plots are obtained via Monte Carlo simulations that evolve $10^8$ independent trajectories to time $T = 20 / \omega_\sys$ and whose initial state is $\tl = \tr = \pi$. The couple of angles at the final time are binned on a grid of size $72 \times 72$. The colour scale is mapped logarithmically and the PDFs are normalised. In the black regions $P_\infty(\tl, \tr)=0$, i.e. none of the trajectories reached that area. 
    Right panels: Flow of $(\tl, \tr)$ under the post-selected no-click dynamics. Red, orange and green dots highlight unstable, saddle and stable points, respectively.}
    \label{fig:gutz_combo}
\end{figure}

The left panels in Fig.\ref{fig:gutz_combo} show a Monte Carlo solution of the stationary distribution $P_\infty(\tl, \tr)\equiv \lim_{t\to\infty} P_t(\tl, \tr)$ for a selection of measurement strengths, $\lambda_1$ and $\lambda_2$.
The three plots represent three different qualitative behaviours of the system.
The transitions between these regimes are sharp and define a phase diagram for the dimer summarized in Fig. \ref{fig:sketch+phase-diagram}.  
For small coupling values, the system is in an ergodic phase, where the measurement strengths are too weak to create forbidden regions.
As a result the whole Hilbert space can be populated at long times.
At larger values of $\lambda_2$ and small enough $\lambda_1$ the system enters a correlated QZ phase where a forbidden region of angles pairs $(\tl, \tr)$ appears close to $\tl = \tr = - \pi$. However $\tl$ and $\tr$ can still assume all values individually.
In other words, a single spin can be found {\it everywhere} on its Bloch sphere, but some combinations of the two angles are forbidden.
This is a consequence of the structure of the two-qubit measurement operator that acts more and more effectively as both spins are close to $-\pi$.  
Finally, when $\lambda_1$ is increased, regardless of the value of $\lambda_2$, the system enters a standard Zeno phase where entire intervals of $\tl$ and $\tr$ are never reached by the dynamics.
We also note that given the symmetry of the unitary and measurement processes, we have $P_\infty(\tl, \tr)=P_\infty(\tr,\tl)$.

From Fig. \ref{fig:gutz_combo} it is clear that the correlated and the standard Zeno regions feature a different topology of the PDF describing the Hilbert space explored by the stochastic evolution.
In the correlated Zeno region, upon mapping on a torus, the allowed portion of the Hilbert space becomes a non-simply connected domain surrounding a non-accessible region.
In the standard Zeno region, local measurements dominates over the correlated ones. The system approaches the standard QZ regime where only a simply connected domain of the physical Hilbert space is explored during the dynamics. 
In the latter case, we observe that 
\begin{equation}
\label{indep_prob}
    P_{\infty}(\tl,\tr)\simeq P_\infty(\tl)P_\infty(\tr),
\end{equation}
reflecting the fact that strong local monitoring decouples the dynamics of the two qubits.
In Eq.\eqref{indep_prob} the equality holds exactly when $\lambda_2=0$.
We conclude by noticing that the structure of the PDF in the correlated Zeno region is a direct consequence of the correlation between the two qubits, which implies $P_{\infty}(\tl,\tr)\neq P_\infty(\tl)P_\infty(\tr)$.

\section{No-click dynamics}
The three distinct behaviours exhibited by the PDF presented above can be understood by studying the no-click evolution of the model given by the $r = 0$ part of Eq. \eqref{eq:dimer-gutz-stochastic-evolution}. 
Upon this post-selection, the dynamics is deterministic and described by the non-Hermitian evolution of the variables $(\tl, \tr)$ as follows 
\begin{equation}
\label{eq:noclick_flow}
    \begin{pmatrix}
        \dot{\theta}_\L(t) \\
        \dot{\theta}_\R(t)
    \end{pmatrix} = -
    \begin{pmatrix}
         \Omega_\L\left(\tl(t), \tr(t)\right) \\
         \Omega_\R\left(\tl(t), \tr(t)\right)
    \end{pmatrix}.
\end{equation}

The right panels of Fig. \ref{fig:gutz_combo} show the flow determined by the differential equation \eqref{eq:noclick_flow} for a selection of values of measurement strengths $\lambda_1$ and $\lambda_2$ \footnote{The flow represents the 2D velocity field of the variables $(\tl, \tr)$ under the no-click dynamics of Eq. \eqref{eq:noclick_flow}. The velocity (tangent to the streamlines) at any point $(\tl, \tr)$ is given by $(\Omega_\L(\tl, \tr), \Omega_\R(\tl, \tr))$.}
In the figure the flow should be thought of as periodic: flow-lines that exit at the bottom and left re-enter from the top and right respectively.
The flow exhibits three qualitatively different behaviours in correspondence of the regions identified previously.
In the ergodic region, the flow has no fixed points.
In the correlated Zeno region the flow develops two fixed points along the diagonal $\tl = \tr$.
One of the two points is unstable (red dot), while the other is a saddle (orange dot).
In the standard Zeno region, the flow develops four fixed points, two of which are again on the diagonal.
Of these two, one is still an unstable point (red dot), while the second is stable (green dot).

The appearance of dynamically forbidden states can be understood from the no-click flow by considering how clicks combine with it.
From a generic point $(\tl, ~\tr)$, clicks can take the system to one of $(\pi, ~\tr)$, $(\tl, ~\pi)$, or $(\pi, ~\pi)$, that is to the right edge, top edge, or right-top corner of the plots, respectively.
In the $t \rightarrow \infty$ limit that we are considering, all trajectories eventually click, independently of the initial conditions.
Without loss of generality, we can thus focus on trajectories with starting points on the top or right edges of the figure.
By inspecting the no-click flow in the right panels of Fig. \ref{fig:gutz_combo}, we can then conclude that in the ergodic region all states are reachable.
In the correlated Zeno region, there is a region close to the bottom-left corner that cannot be reached by trajectories that have their starting point on the top and right edges;
In standard Zeno region then entire intervals of $\tl$ and $\tr$ are eventually depleted by clicks that take the system to the state $(\pi, ~\pi)$.

\begin{figure}[t]
    \centering
    \includegraphics[scale=0.52]{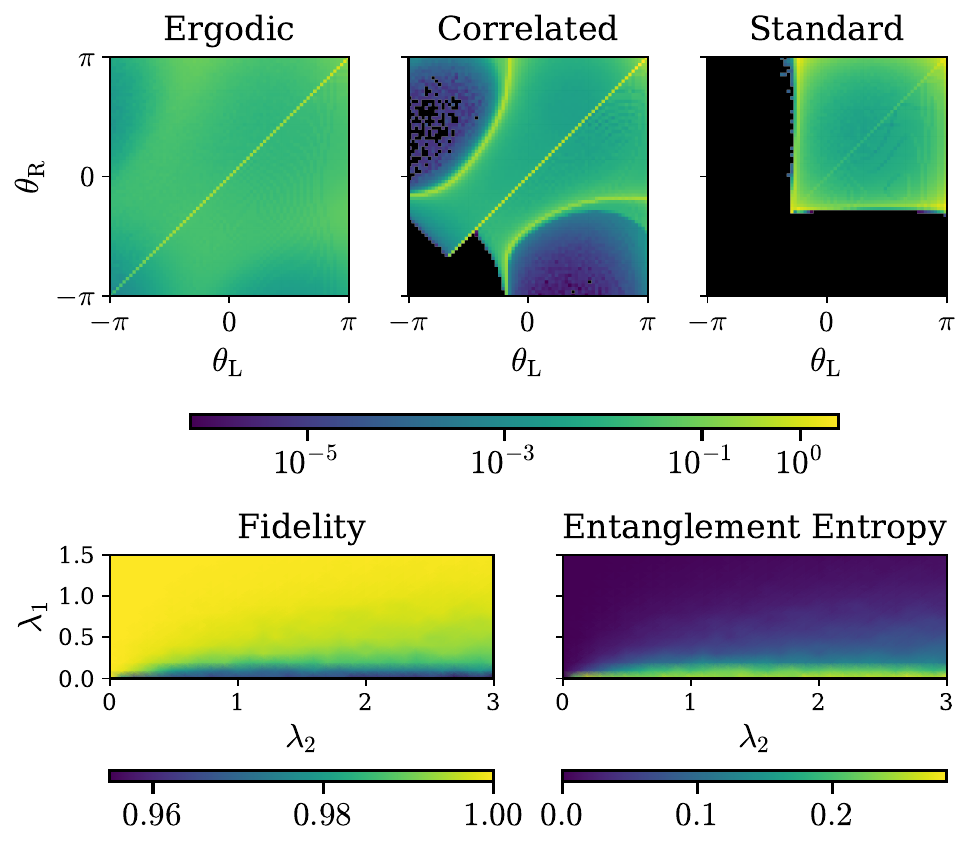}
    \caption{Top panels: Comparison of exact and Gutzwiller stationary PDF $P_\infty(\tl, \tr)$ in the three phases. Exact simulations are shown above the anti-diagonal, simulation in the Gutzwiller approximation (taken from Fig. \ref{fig:gutz_combo}) below. For the exact simulations $N_{\rm traj}>10^6$.
    Bottom panels: Stationary average fidelity $\overline{F}$ and stationary average entanglement entropy $\overline{S}_{E}$. Here $N_{\rm traj}>10^3$.
    In all the panels, the parameters of the numerics are as in Fig. \ref{fig:gutz_combo}.
    }
    \label{fig:dimer-full-vs-gutz}
\end{figure}

Summarizing, the three distinct regimes of the two-qubit system under varying measurement strengths $\lambda_1$ and $\lambda_2$ are in one-to-one correspondence with the structure of the flow of the no-click evolution.
The boundaries between the different regimes depicted in Fig. \ref{fig:sketch+phase-diagram} can be thus obtained by studying the structure of the PDF or, equivalently, computing the fixed points of the $(\tl(t), \tr(t))$ flow under the non-Hermitian evolution.

Particularly, the boundary between the ergodic and the Zeno regimes is determined by the condition for the appearance of diagonal fixed points through a saddle-node bifurcation, while the boundary between the correlated and standard Zeno regimes is set by the change in stability of the fixed point in $\tl=\tr\in[-\pi/2,0]$.
A full semi analytical treatment to determine these boundaries is presented in Appendix~\ref{app:phase_diag}.
Notably, the latter boundary can be written as
\begin{equation}
\mathcal{B}(\lambda_1,\lambda_2)
=
[\lambda_1(\lambda_1+\lambda_2)]^3
-
\left(\lambda_1+\frac{\lambda_2}{2}\right)^4
=0,
\label{eq:Hboundary_main}
\end{equation}
which separates the stable ($\mathcal{B}(\lambda_1,\lambda_2)>0$) and saddle ($\mathcal{B}(\lambda_1,\lambda_2)<0$) character of the fixed point.
Interestingly, the critical single-qubit measurement rate $\lambda_1^{(\rm{c})}$ as a function of $\lambda_2$ (obtained by solving $\mathcal{B}(\lambda_1^{(\rm{c})},\lambda_2)=0$) determines the phase boundary and is not monotonous as a function $\lambda_2$. 
Specifically we get $\lambda_1^{(\rm{c})}= 1$ for $\lambda_2=0$ and a decreasing function with a minimum at $\lambda_2 = 8/(3\sqrt{3})\approx 1.54$ (where $\lambda_1^{(\rm{c})}=4/(3\sqrt{3})\approx 0.77$). 
Above such value $\lambda_1^{(\rm{c})}$ increases with asymptotic behavior $\lambda_1^{(\rm{c})}\sim(\lambda_2/16)^{1/3}$ for $\lambda_2/\lambda_1\gg1$.
This behaviour is not evident in the parameter range we chose for the phase diagram in Fig. \ref{fig:sketch+phase-diagram} where the phase boundary appears rather flat (see Appendix \ref{app:phase_diag}). 

We close this section stressing that the topologically distinct regimes are a key feature of the quantum-jump dynamics.
Different measurement schemes, such as quantum state diffusion, lead instead to the disappearance of the forbidden region, as discussed in Appendix~\ref{sec:stoch-appendix}.

\section{Comparison with full dynamics}
The main difference between the Gutzwiller and full dynamics is that the former is not able to capture quantum entanglement between the two qubits. The disagreement between the results is a measure of the relevance of quantum correlations in the dynamics of the system.

The top panels in Fig. \ref{fig:dimer-full-vs-gutz} show that the stationary PDF for the two angles, $\tl$ and $\tr$, computed by simulating the full dynamics of the system, exhibit the same qualitative behaviour, with three distinct regimes as described previously in the Gutzwiller approximation \footnote{Given that the full wavefunction is in general not factorizable one has to compute the on-site reduced density matrix for the $\L/\R$ qubit and exploit the Bloch sphere representation which gives directly access to $\theta^{(i)}_{\L/\R}(t)$ and thus compute the PDF.}.
In order to get further insights about the accuracy of the  approximation in the different regimes we computed fidelity between the exact trajectories and the Gutzwiller ones and the entanglement entropy of the former, respectively 
\be
\begin{split}
\overline{F} &=\frac{1}{N_\traj} \sum_{i=1}^{N_\traj} 
\left|
\langle\tl^{(i)},\tr^{(i)}|\Psi^{(i)}_{\rm EX}\rangle
\right|^2, \\ 
\overline{S}_{E} &= \frac{1}{N_\traj}\sum_{i=1}^{N_\traj} - \Tr{\rho^{(i)}_{\rm L, EX} \log_2 \rho^{(i)}_{\rm L, EX}}
\end{split}
\ee
where $|\Psi^{(i)}_{\rm EX}\rangle$ is the exact state along the $i$-th trajectory, $\rho^{(i)}_{\rm L, EX}={\rm Tr}_{\rm R}[|\Psi^{(i)}_{\rm EX}\rangle\langle\Psi^{(i)}_{\rm EX}|]$ is the corresponding reduced density matrix of the L qubit and $|\tl^{(i)},\tr^{(i)}\rangle$ is its closest Gutzwiller state.

Unsurprisingly, the quantitative differences that can be observed between the full and Gutzwiller PDFs are greater at high values of the $\lambda_2 / \lambda_1$ ratio. This can be explained by observing that the only entangling coupling between the two qubits is given by the two-body measurement part of the non-Hermitian Hamiltonian \eqref{nonHermHam}, whose strength is set by $\lambda_2$.
Interestingly, also at large values of $\lambda_2$, moderate values of $\lambda_1$ are sufficient to strongly suppress quantum entanglement of the trajectories.

\section{Conclusions}
In this work, we have proposed a theory to explain the emergence of the QZ effect in a monitored qubit dimer. Our analysis has led to the identification of two qualitatively distinct QZ regimes, each characterized by a different topology in the probability density function (PDF). This behavior was further linked to the underlying topology of the no-click flow in the system.

Building on these foundational results, our theory can be extended to chains made of $N$ qubits. In general this allows the study of the interplay between on-site unitary dynamics and $k-$qubit measurements described by $M_i^k = \sqrt{\gamma_k dt}\prod_{i=1}^{i+k}n_i$ (where $1 \le k \le N$ and $i$ lables the qubits).
A promising direction for future research would be thus the exploration of how the topology of allowed and forbidden regions varies as a function of different multi-site measurement protocols via the generalization of the stochastic Gutzwiller ansatz.
When $\gamma_2$ dominates over the other scales we expect that the probability distribution function of all the pairs of adiaject spins will exhibith a topology similar to the one showed in this paper.
When a given $\gamma_k$ with $k\geq 3$ dominates we do expect signatures in the $k-$qubit probability distribution function (constructed analogously, considering the angles $(\theta_i, \theta_{i+1}, \dots, \theta_{i+k-1})$) will display topological transitions in their structure.

{\it Acknowledgements -- }
We thank Rosario Fazio, Marcello Dalmonte, Philipp Hauke, Alessandro Roggero and Oded Zilberberg for inspiring discussions.  
AB acknowledges financial support
from the Provincia Autonoma di Trento and by the European Union —
NextGeneration EU, within PRIN 2022, PNRR M4C2, Project TANQU 2022FLSPAJ [CUP B53D23005130006]. AR and GC acknowledge support by EPSRC via  Grant No. EP/W524438/1.

{\it Data availability-- }
The data that support the findings of this article are openly available in \cite{severino_2025_15260513}.

\appendix

\section{Measurement protocol and its physical model}
\label{app:phys_mod}

The system considered in this work is composed of two qubits, referred to as left ($\L$) and right qubit ($\R$).
Their evolution is the combined result of the unitary evolution under the non-interacting Hamiltonian
\begin{equation}
    H_\sys=
    \omega_\sys \left( \paulix_{\L} + \paulix_{\R} \right),
\end{equation}
where $\omega_\sys > 0$, and continuous monitoring via a generalised measurement that is performed on them at time intervals $dt \ll 1 / \omega_\sys$.
This protocol is represented schematically in figure \ref{fig:sketch+phase-diagram} in the main text.

The measurement that we consider has four possible outcomes, labelled $r = 0, \dots, 3$.
Outcomes $r=1$ and $r=2$ account for local weak measurements of the $\L$ and $\R$ qubit populations respectively.
The corresponding back-actions on the system is given by the Kraus operators 
\begin{equation} \label{eq:dimer-effects-lr}
    \begin{split}
        &M_1 = \sqrt{p_1} n_\L \\
        &M_2 = \sqrt{p_1} n_\R,
    \end{split}
\end{equation}
where $n_\LR = (\pauliz_\LR + \qeye) / 2$ and $p_1 \in [0, 1]$ controls the local measurement strength (which is the same for both qubits).
The $r = 3$ outcome accounts for the joint measurement of the $\L$ and $\R$ qubit populations.
Its back-action is
\begin{equation} \label{eq:dimer-effects-b}
    M_3
    = \sqrt{p_2} n_\L n_\R
\end{equation}
where $p_2 \in [0, 1]$ controls the two-body measurement strength.
Finally, the $r=0$ outcome back-action is
\begin{equation} \label{eq:dimer-effects-0}
    M_0
    = \sqrt{\qeye - M_\L^2 - M_\R^2 - M_\B^2}.
\end{equation}
In the literature, and for reasons that will become clear in what follows, the $0$ outcome is referred to as ``no-click'' while the $r = 1, \dots, 3$ outcomes are all referred to as ``clicks''.

As usual, the probability of each of the outcomes is given by
\begin{equation}
    p_r
    = \bra{\psi} \adj{M}_r M_r \ket{\psi}
    \label{eq:dimer-readout-probability}
\end{equation}
for $r = 0, \dots, 3$.

A physical model of the measurement consists of three auxiliary qubits, $\detec_\L$, $\detec_\R$ and $\detec_\B$, that interact with the two qubits comprising the dimer.
The three detector qubits are coupled to the dimer via the following Hamiltonians
\begin{equation}
    \begin{split}
        &H_{\sys-\detec_\LR} = \frac{J_1}{2} (\qeye - \pauliz_\LR) \pauliy_{\detec_\LR}, \\
        &H_{\sys-\detec_\B} = \frac{J_2}{4} (\qeye - \pauliz_{\L}) (\qeye - \pauliz_{\R}) \pauliy_\B,
    \end{split}
\end{equation}
The measurement procedure is repeated at short time intervals $dt$.
At the beginning of each time interval, the three auxiliary qubits are initialised in their $\ket{0}$ state, then they are left to interact coherently with the system, and are subsequently measured projectively in their $\left\{\ket{0}, \ket{1}\right\}$ basis at the end of the time interval.
The initialisation and readout are assumed to last a negligible time.

Therefore, in the time interval between successive measurements, the system evolves under the total Hamiltonian given by the sum of the system's plus the interaction with the detectors
\begin{equation}
    H
    = H_\sys + H_{\sys-\detec_\L} + H_{\sys-\detec_\R} + H_{\sys-\detec_\B}.
\end{equation}
While the effect of the measurement on the system is obtained by computing the following back-actions
\begin{equation} \label{eq:dimer-povm-effects-definition}
    M_{(i, j, k)}
    = \bra{i_{\detec_\L}, j_{\detec_\R}, k_{\detec_\B}} e^{-i H_{\sys-\detec} dt} \ket{0_{\detec_\L}, 0_{\detec_\R}, 0_{\detec_\B}},
\end{equation}
where $i, j, k = 0, 1$ index the possible outcomes of the measurements on the ancillary qubits.

Since the system-detector Hamiltonians all commute with each other, the evolution operator in the equation above can be factorised
\begin{equation}
    e^{-i H_{\sys-\detec} dt}
    = e^{-i H_{\sys-\detec_\L} dt} e^{-i H_{\sys-\detec_\R} dt} e^{-i H_{\sys-\detec_\B} dt},
\end{equation}
hence, the expression for the back-actions can also be factorised into contributions that involve only one detector each
\begin{equation} \label{eq:dimer-effect-factorisation}
    M_{(i, j, k)}
    = M_{\detec_\L, i} M_{\detec_\R, j} M_{\detec_\B, k},
\end{equation}
with
\begin{equation} \label{eq:dimer-single-detector-povm-definition}
    M_{\detec_p, i}
    = \bra{i_{\detec_p}} e^{-i H_{\sys-\detec_p} dt} \ket{0_{\detec_p}},
\end{equation}
where $p$ labels the detector, $p = \L, \R, \B$.

Straightforward manipulations of the system-detector evolution operators lead to
\begin{equation}
    \begin{split}
        e^{-i H_{\sys-\detec_\LR} dt}
        = &\Bigl[ \ket{0_\LR}\bra{0_\LR} + \ket{1_\LR}\bra{1_\LR} \cos(J_1 dt) \Bigr] \qeye[\detec_\LR] \\
        - &\Bigl[i \ket{1_\LR}\bra{1_\LR} \sin(J_1 dt)\Bigr] \pauliy_{\detec_\LR}
    \end{split}
\end{equation}
and
\begin{equation}
    \begin{split}
        e^{-i H_{\sys-\detec_\B} dt}
        = &\Bigl[ \ket{0_\L 0_\R}\bra{0_\L 0_\R} + \ket{1_\L 1_\R}\bra{1_\L 1_\R} \cos(J_2 dt) \Bigr] \qeye[\detec_\B] \\
        - &\Bigl[i \ket{1_\L 1_\R}\bra{1_\L 1_\R} \sin(J_2 dt)\Bigr] \pauliy_{\detec_\B}.    
    \end{split}
\end{equation}
Performing the inner product in equation \ref{eq:dimer-single-detector-povm-definition}, and extracting the leading order correction in the small time interval $dt$, in the continuous monitoring limit where $J_i^2 dt = \gamma_i = \text{const}$ (i.e. $J_i = \sqrt{\gamma_i / dt}$), yields
\begin{equation}
    \begin{split}
        M_{\detec_\L, 0}
        &= \diag{1, 1, \cos(J_1 dt), \cos(J_1 dt)} \\
        &\simeq \diag{1, 1, 1-\frac{\gamma_1}{2} dt, 1-\frac{\gamma_1}{2} dt}, \\
        M_{\detec_\L, 1}
        &= \diag{0, 0, \sin(J_1 dt), \sin(J_1 dt)} \\
        &\simeq \diag{0, 0, \sqrt{\gamma_1 dt}, \sqrt{\gamma_1 dt}},
    \end{split}
\end{equation}
\begin{equation}
    \begin{split}
        M_{\detec_\R, 0}
        &= \diag{1, \cos(J_1 dt), 1, \cos(J_1 dt)} \\
        &\simeq \diag{1, 1-\frac{\gamma_1}{2} dt, 1, 1-\frac{\gamma_1}{2} dt}, \\
        M_{\detec_\R, 1}
        &= \diag{0, \sin(J_1 dt), 0, \sin(J_1 dt)} \\
        &\simeq \diag{0, \sqrt{\gamma_1 dt}, 0, \sqrt{\gamma_1 dt}}
    \end{split}
\end{equation}
\begin{equation}
    \begin{split}
        M_{\detec_\B, 0}
        &= \diag{1, 1, 1, \cos(J_2 dt)} \\
        &\simeq \diag{1, 1, 1, 1-\frac{\gamma_2}{2} dt}, \\
        M_{\detec_\B, 1}
        &= \diag{0, 0, 0, \sin(J_2 dt)} \\
        &\simeq \diag{0, 0, 0, \sqrt{\gamma_2 dt}}.
    \end{split}
\end{equation}

The total back-action matrices are obtained from these via equation \ref{eq:dimer-effect-factorisation} by multiplying the effects above for all combinations of $i, j, k = 0, 1$
However, of all eight possible combinations, the four with more than one detector click can be discarded.
This can be seen by considering the probabilities associated with the possible outcomes, which for some state of the system $\ket{\psi}$ are given by
\begin{equation}
    p_{(i, j, k)}
    = \bra{\psi} \adj{M_{(i, j, k)}} M_{(i, j, k)} \ket{\psi}
\end{equation}
Using the expressions of the single detector effects above, it is straightforward to see that their respective leading order in $dt$ is
\begin{equation}
    \begin{split}
        &p_{(0, 0, 0)} = O(1), \\
        &p_{(1, 0, 0)}, ~p_{(0, 1, 0)}, p_{(0, 0, 1)} = O(dt), \\
        &p_{(1, 1, 0)}, ~p_{(1, 0, 1)}, p_{(0, 1, 1)} = O(dt^2), \\
        &p_{(1, 1, 1)} = O(dt^3),
    \end{split}
\end{equation}
hence, the outcomes with more than 1 detector click have a negligible probability of happening and can be discarded.
The remaining effects are (up to order $O(dt^2)$)

\begin{widetext}
\begin{equation}
\begin{aligned}
M_\L &= M_{(1,0,0)} = \sqrt{\gamma_1 dt}
\begin{pmatrix}
0 & 0 & 0 & 0 \\
0 & 0 & 0 & 0 \\
0 & 0 & 1 & 0 \\
0 & 0 & 0 & 1
\end{pmatrix}, \\
M_\R &= M_{(0,1,0)} = \sqrt{\gamma_1 dt}
\begin{pmatrix}
0 & 0 & 0 & 0 \\
0 & 1 & 0 & 0 \\
0 & 0 & 0 & 0 \\
0 & 0 & 0 & 1
\end{pmatrix}, \\
M_\B &= M_{(0,0,1)} = \sqrt{\gamma_2 dt}
\begin{pmatrix}
0 & 0 & 0 & 0 \\
0 & 0 & 0 & 0 \\
0 & 0 & 0 & 0 \\
0 & 0 & 0 & 1
\end{pmatrix}, \\
M_0 &= M_{(0,0,0)} =
\begin{pmatrix}
1 & 0 & 0 & 0 \\
0 & 1-\frac{\gamma_1}{2}dt & 0 & 0 \\
0 & 0 & 1-\frac{\gamma_1}{2}dt & 0 \\
0 & 0 & 0 & 1-\frac{1}{2}(2\gamma_1+\gamma_2)dt
\end{pmatrix}.
\end{aligned}
\label{eq:appendix-effects-final}
\end{equation}
\end{widetext}
i.e. those in equations \ref{eq:dimer-effects-lr}, \ref{eq:dimer-effects-b} and \ref{eq:dimer-effects-0} with the measurement strength parameters $p_{1, 2}$ now expressed as $p_{1, 2} = \gamma_{1, 2} dt$.

\section{Derivation of the Gutzwiller stochastic dynamics}
\label{sec:dime-gutz-derivation}
\begin{figure}[t]
    \centering
    \includegraphics[scale=0.25]{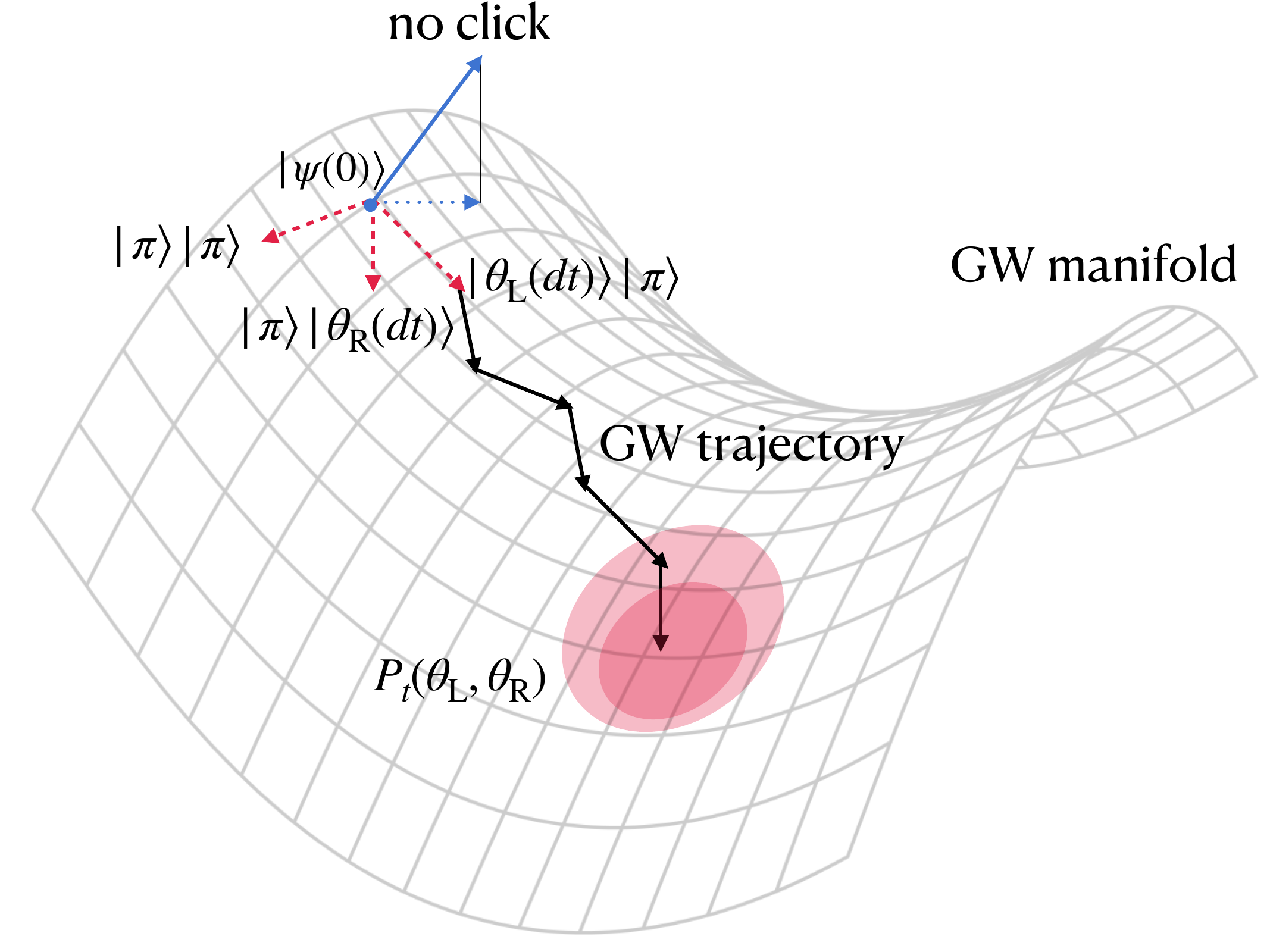}
    \caption{Pictorial representation of a quantum Gutzwiller trajectory. 
    Starting from a product state $\ket{\psi(0)}=\ket{\tl(0)}\ket{\tr(0)}$ the measurement outcomes $r=1,2,3$ (dashed arrows) leave the state within the Gutzwiller manifold parametrized by the pair of angles $(\tl,\tr)$ while the no-click outcome $r=0$ lead to a non-factorizable state lying outside the manifold (solid arrow). In the latter case the state is projected back according to the prescription \eqref{eq:dimer-gutz-stochastic-evolution}. The dynamics thus lead to a trajectory that stochastically evolves within the Gutzwiller manifold and generate the PDF $P_t(\tl,\tr).$ 
    }
    \label{fig:sketch_GW}
\end{figure}

In this appendix the Gutzwiller stochastic equation (\ref{eq:dimer-gutz-stochastic-evolution}) in the main text is derived. The dynamics is represented pictorially in Fig.\ref{fig:sketch_GW}.
The focus is on the no-click outcome of the measurement procedure, as the other cases lead to projections of the system state (clicks).

The continuous part of the evolution is generated by the effective, non-Hermitian Hamiltonian
\begin{equation}
    H_\eff
    = H - \frac{i}{2}  \sum_i \left( \adj{L_i} L_i + \ev{\adj{L_i} L_i} \right).
\end{equation}

In the case at hand the system Hamiltonian is
\begin{equation}
    H_\sys
    = \omega_\sys \left(\paulix_{\L} + \paulix_{\R}\right),
\end{equation}
and the jump operators
\begin{equation}
    \begin{split}
        L_\L &= \sqrt{\gamma_1} n_\L, \\
        L_\R &= \sqrt{\gamma_1} n_\R, \\
        L_\B &= \sqrt{\gamma_2} n_\L n_\R,
    \end{split}
\end{equation}
where the $n_\L$, $n_\R$ are the projectors on the up state of the left, right, and both qubits respectively, $n_\LR = \ket{1_\LR} \bra{1_\LR}$, and $\gamma_1$ and $\gamma_2$ are parameters related to the strength of the local and non-local measurements respectively.

We want to find the (no-click) equation of motion of the system in the Gutzwiller approximation, that is assuming that the wavefunction of the system remains a product of two one-qubit wavefunctions at all times
\begin{equation}
    \ket{\psi(t)}
    = \ket{\psi_\L(t)} \ket{\psi_\R(t)}.
\end{equation}
Equations for the evolution of the $\ket{\psi_\L(t)}$  and $\ket{\psi_\R(t)}$ states can be found by plugging the Gutzwiller ansatz into the action
\begin{equation}
    S
    = \int_{t_i}^{t_f} \bra{\psi(t)} \left( -i \frac{\diff}{\diff t} + H_\eff \right) \ket{\psi(t)} \diff t,
\end{equation}
and, varying $\bra{\psi_\L(t)}$ and $\bra{\psi_\R(t)}$ respectively.
For the $\L$ qubit, this yields

\begin{equation}
\label{eq:dimer-gutz-schroedinger}
\begin{split}
i \frac{\diff}{\diff t} \ket{\psi_\L(t)}
={}& \Biggl[
\omega_\sys \paulix_\L
- i \frac{\gamma_1}{2} \bigl( n_\L - \ev{n_\L} \bigr) \\
&\qquad
- i \frac{\gamma_2}{2}
\bigl( \ev{n_\R} n_\L - \ev{n_\L}\ev{n_\R} \bigr)
\Biggr]\ket{\psi_\L(t)} .
\end{split}
\end{equation}
while the equation for the $\R$ qubit can be obtained from the above by exchanging the role of $\L$ and $\R$.
The two equations represent the evolution of the individual qubits coupled via mean-field terms.

Because of the specific choice of system Hamiltonian $H_\sys$ and jump operators $\{n_i\}$, the dynamics of the two qubits is confined to the $y-z$ plane of their respective Bloch spheres, when these are initialised on the same plane.
This allows one to parametrise the state of each system with a single angle.
By extending the usual $\theta$ angle of the Block sphere also to the $(-\pi, 0)$ interval, one can always write
\begin{equation}
    \ket{\psi_q(t)}
    = \cos(\theta_q(t)/2) \ket{0_q} + i \sin(\theta_q(t)/2) \ket{1_q},
\end{equation}
where $q = \LR$.

By plugging this form of the states into equation \ref{eq:dimer-gutz-schroedinger}, and projecting it on the $\ket{0_\LR}$ state, it is straightforward to obtain the the following coupled dynamical equations
\begin{equation} \label{eq:dimer-gutz-thetas-dynamics}
    \begin{cases}
        \dot{\tl}(t)
        = 2 \omega_\sys \left( 1 + \left(\lambda_1 + \lambda_2 \trig[2]{\sin}{\tr / 2} \right) \trig{\sin}{\tl} \right) \\
        \dot{\tr}(t)
        = 2 \omega_\sys \left( 1 + \left(\lambda_1 + \lambda_2 \trig[2]{\sin}{\tl / 2} \right) \trig{\sin}{\tr} \right)
    \end{cases}
\end{equation}
where $\lambda_{1,2} = \gamma_{1, 2} / (4 \omega_\sys)$.

Supplemented with the click outcomes, this becomes equation \ref{eq:dimer-gutz-stochastic-evolution}.

\section{Gutzwiller master equation}
\label{app:gw_master_eq}
If one considers an ensemble of systems, this can be described by the probability density of finding one of these in a given state, at a given time.
Within the Gutzwiller ansatz, this can be written as a function of the two parameters $\tl$ and $\tr$ (as well as time $t$), $P_t(\tl, \tr)$. $P_t(\tl, \tr)$ represents the joint probability of finding the left qubit in state $\ket{\tl}$ and the right qubit in state $\ket{\tr}$ (at time $t$).

Given an ensemble of trajectories, one can write the following formal definition
\begin{equation}
    P_t(\tl, \tr)
    = \frac{1}{N_\traj} \sum_{i=1}^{N_\traj}
        \delta\left( \tl^{(i)}(t)  - \tl \right)
        \delta\left( \tr^{(i)}(t)  - \tr \right)
\end{equation}
where $\bigl( \tl^{(i)}(t), ~\tr^{(i)}(t) \bigr)$ is the state of the system at time $t$ along the $i$-th trajectory and  $N_{\rm traj}$ is the number of trajectories in the ensemble.
From the definition above it is clear that $P_t(\tl, \tr)$ only contains classical correlations between the two qubits.

From the Gutzwiller stochastic equation (Eq. \ref{eq:dimer-gutz-stochastic-evolution}) and the readout probabilities (Eq. \ref{eq:dimer-gutz-readout-probabilities}), it is straightforward to derive the following master equation for the time evolution of the ensemble probability distribution

\begin{widetext}
\begin{equation}
\label{eq:dimer-fokker-planck}
\begin{split}
\frac{\diff P_t(\tl,\tr)}{\diff t} ={}&
\partial_{\tl}\!\left[\Omega_\L(\tl,\tr)\,P_t(\tl,\tr)\right]
+\partial_{\tr}\!\left[\Omega_\R(\tl,\tr)\,P_t(\tl,\tr)\right]
\\
&-\gamma_1\!\left(\trig[2]{\sin}{\tl/2}+\trig[2]{\sin}{\tr/2}\right)P_t(\tl,\tr)
-\gamma_2\,\trig[2]{\sin}{\tl/2}\trig[2]{\sin}{\tr/2}\,P_t(\tl,\tr)
\\
&+\gamma_1\,\delta(\tl-\pi)
\int_{-\pi}^{+\pi}\diff\tl'\,
\trig[2]{\sin}{\tl'/2}\,P_t(\tl',\tr)
\\
&+\gamma_1\,\delta(\tr-\pi)
\int_{-\pi}^{+\pi}\diff\tr'\,
\trig[2]{\sin}{\tr'/2}\,P_t(\tl,\tr')
\\
&+\gamma_2\,\delta(\tl-\pi)\delta(\tr-\pi)
\int_{-\pi}^{+\pi}\diff\tl'
\int_{-\pi}^{+\pi}\diff\tr'\,
\trig[2]{\sin}{\tl'/2}\trig[2]{\sin}{\tr'/2}\,
P_t(\tl',\tr').
\end{split}
\end{equation}
\end{widetext}

The terms of this equation can be interpreted as follows.
The partial derivatives represent the drift due to the no-click evolution.
The following two terms account for the depletion of the probability distribution function due to clicks.
While the final, integral bits describe the accumulation of clicks onto the measurement outcome states.

{\section{Phase diagram from Gutzwiller Ansatz}
\label{app:phase_diag}
Under the Gutzwiller approximation, after post-selection the dynamics is governed by a non-Hermitian flow for the Bloch angles $(\theta_L,\theta_R)$:
\begin{equation}
\begin{pmatrix}
\dot\theta_L \\[4pt]
\dot\theta_R
\end{pmatrix}
=
-
\begin{pmatrix}
\Omega_L(\theta_L,\theta_R) \\[4pt]
\Omega_R(\theta_L,\theta_R)
\end{pmatrix},
\end{equation}
with
\begin{align*}
\Omega_L(\theta_L,\theta_R)
&=
2ws\left[1+\left(\lambda_1+\lambda_2\sin^2\frac{\theta_R}{2}\right)\sin\theta_L\right], \\
\Omega_R(\theta_L,\theta_R)
&=
2ws\left[1+\left(\lambda_1+\lambda_2\sin^2\frac{\theta_L}{2}\right)\sin\theta_R\right].
\end{align*}
Fixed point in the dynamics implies, $\Omega_{L,R}=0$ which restricts them along diagonal $\theta_L=\theta_R=\theta$. In this case $\Omega(\theta)=0$ implies 
\begin{equation}\label{fx_pt}
    \frac{1}{\sin{\theta}}+\left(\lambda_1+\lambda_2\sin^2\frac{\theta}{2}\right) = 0
\end{equation}
Since $\lambda_{1,2}>0$, we can get fixed points for $\theta\in [-\pi,0)$.

We rewrite the fixed-point Eq.~\ref{fx_pt} in terms of
\(r=|\sin\theta|\in(0,1]\) and
\(s=\mathrm{sign}(\cos\theta)=\pm1\),
\begin{equation}
g_s(r)= -\frac{1}{r}
+\Bigl(\lambda_1+\frac{\lambda_2}{2}\Bigr)
- s\,\frac{\lambda_2}{2}\sqrt{1-r^2}.
\label{eq:gs_def}
\end{equation}


For the \(s=-1\), \(g_-(r)\) has a single maximum at some \(r=r_0\in(0,1)\)
determined by
\begin{equation}
g_-'(r_0)=0.
\label{eq:r0_def}
\end{equation}

which gives
\begin{equation}
\frac{1}{r_0^2}-\frac{\lambda_2}{2}\frac{r_0}{\sqrt{1-r_0^2}}=0.
\label{eq:r0_eq}
\end{equation}

This equation has a unique positive root
\begin{equation}
r_0(\lambda_2)=
\frac{1}{\sqrt{3}}\,
\sqrt{
\frac{6^{1/3}\lambda_2^{2}
\left(
\frac{9+\sqrt{48+81\lambda_2^{2}}}{\lambda_2^{3}}
\right)^{2/3}
-2\,(6^{2/3})}{
\lambda_2^{2}
\left(
\frac{9+\sqrt{48+81\lambda_2^{2}}}{\lambda_2^{3}}
\right)^{1/3}
}
}.
\end{equation}
This determines the position of the maximum of
\(g_-(r)\).
Having determined the location \(r_0(\lambda_2)\) of the maximum of \(g_-(r)\),
we evaluate the maximal value
\begin{equation}
g_-(r_0(\lambda_2))= F(\lambda_1,\lambda_2)=
-\frac{1}{r_0(\lambda_2)}
+\Bigl(\lambda_1+\frac{\lambda_2}{2}\Bigr)
+\frac{\lambda_2}{2}\sqrt{1-r_0(\lambda_2)^2}.
\label{eq:gmax}
\end{equation}
Since $g_-(r)$ is unimodal on $r\in(0,1]$, the number of solutions of
the fixed-point equation $g_-(r)=0$ is completely determined by the sign
of $F(\lambda_1,\lambda_2)$: $F>0$ yields two fixed points,
$F=0$ a single (saddle–node) fixed point, and $F<0$ no fixed point. These conditions defines the phase boundaries in the \((\lambda_1,\lambda_2)\)
plane, clearly separating regions where fixed points exist from those where they do not with, as shown
in Fig.~\ref{fig:fixpt_phase}.
\begin{figure}[!h]
\centering
\includegraphics[scale=0.45]{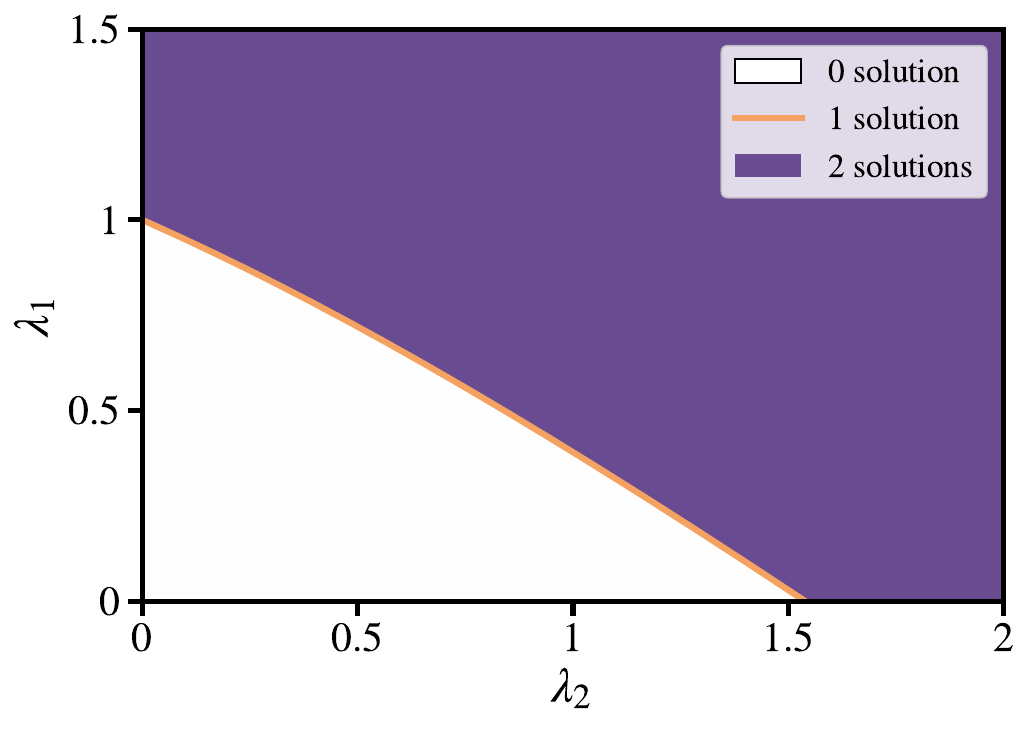}
\caption{Phase diagram in the \((\lambda_1,\lambda_2)\) plane obtained for two, one, or no fixed points.}
\label{fig:fixpt_phase}
\end{figure}

We now determine the nature of the fixed points from the Jacobian. For general $(\theta_L,\theta_R)$ the Jacobian reads
\begin{equation}
J(\theta_L,\theta_R)
=
-
\begin{pmatrix}
\partial_{\theta_L}\Omega_L & \partial_{\theta_R}\Omega_L\\
\partial_{\theta_L}\Omega_R & \partial_{\theta_R}\Omega_R
\end{pmatrix}.
\end{equation}

On the diagonal this simplifies to
\begin{equation}
J(\theta)
=
-2ws
\begin{pmatrix}
(\lambda_1+\lambda_2\sin^2\frac{\theta}{2})\cos\theta &
\frac{\lambda_2}{2}\sin^2\theta\\[6pt]
\frac{\lambda_2}{2}\sin^2\theta &
(\lambda_1+\lambda_2\sin^2\frac{\theta}{2})\cos\theta
\end{pmatrix}
\end{equation}
with eigenvalues
\begin{equation}
\mu_\pm(\theta)
=
-2ws\!\left[
(\lambda_1+\lambda_2\sin^2\frac{\theta}{2})\cos\theta
\pm
\frac{\lambda_2}{2}\sin^2\theta
\right],
\end{equation}
and corresponding eigenvectors $(1,\pm1)/\sqrt2$.

The stability of a fixed point $\theta^*$ is therefore determined by the
signs of $\mu_\pm$: it is a stable fixed point if $\mu_\pm<0$, a saddle if
$\mu_+\mu_-<0$, and unstable fixed point if $\mu_\pm>0$.

Using the fixed-point relation
the eigenvalues can be rewritten as
\begin{align}
\mu_- &= -2ws\left(\frac{s\sqrt{1-r^2}}{r}-\frac{\lambda_2}{2}r^2\right),\\
\mu_+ &= -2ws\left(\frac{s\sqrt{1-r^2}}{r}+\frac{\lambda_2}{2}r^2\right).
\end{align}

For the $s=+1$, ($\theta\in[-\pi/2,0]$) one has $\mu_+<0$ for all
$\lambda_2$, while $\mu_-<0$ iff $\lambda_2<2\sqrt{1-r^2}/r^3$.
An attractor therefore exists only in the parameter regime
\begin{equation*}
\lambda_2<\frac{2\sqrt{1-r^2}}{r^3}.
\label{eq:attractor_cond}
\end{equation*}
Thus, for $s=+1$, the boundary of the stable to the saddle fixed point region follows from
$\lambda_2=2\sqrt{1-r^2}/r^3$. Substituting this into the fixed-point
equation yields the parametric curve
\begin{equation*}
\lambda_1=\frac{1-\sqrt{1-r^2}}{r^3}.
\end{equation*}
Eliminating $r$ gives the explicit phase boundary
\begin{equation}
\mathcal{B}(\lambda_1,\lambda_2)
=
[\lambda_1(\lambda_1+\lambda_2)]^3
-
\left(\lambda_1+\frac{\lambda_2}{2}\right)^4
=0.
\end{equation}

Consequently, $\mathcal{B}>0$ corresponds to the stable fixed point and $\mathcal{B}<0$ correspond to the
saddle fixed point.

For $s=-1$ ($\theta\in[-\pi,-\pi/2]$) we get $\mu_- > 0$ for all $\lambda_2$. 
Furthermore, $\mu_+ > 0$ for all values of $\lambda_2$, for which solution exist. Hence the $s=-1$ fixed point is always repulsive.
These regions are shown in Fig.~\ref{fig:nat_fixpts}.

\begin{figure}[!h]
\centering
\includegraphics[scale = 0.45]{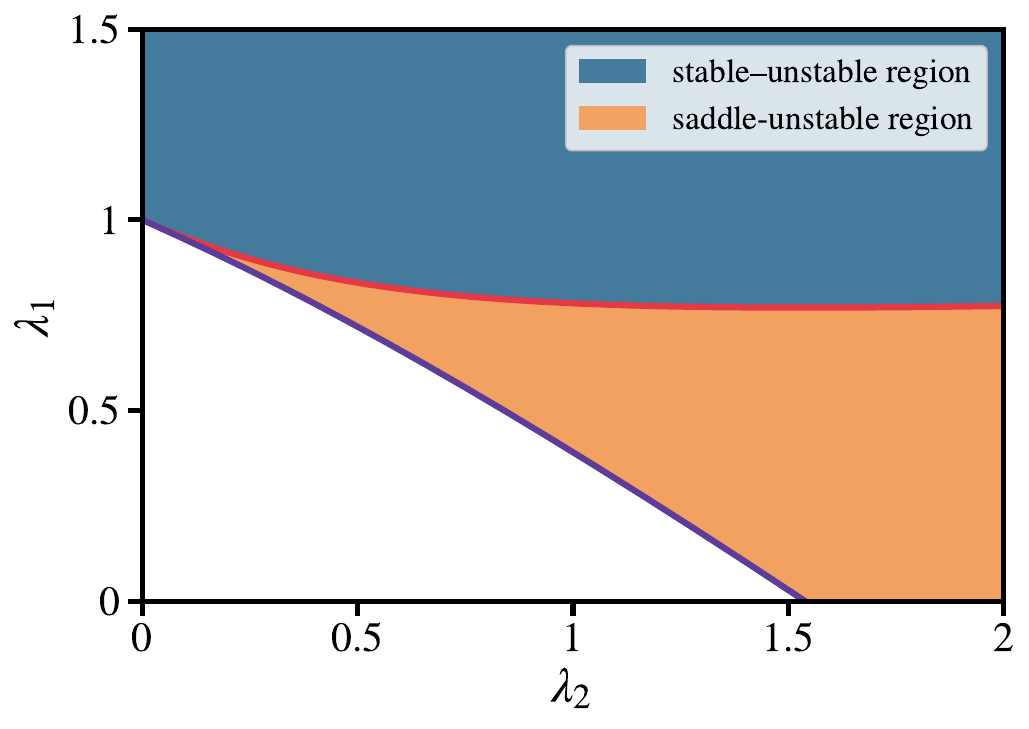}
\caption{Phase diagram in the \((\lambda_1,\lambda_2)\) plane obtained for nature of fixed points.}
\label{fig:nat_fixpts}
\end{figure}
We can further comment on the structure of the boundary \( \mathcal{B}(\lambda_1,\lambda_2) \). Although, within the parameter window displayed in Fig.~\ref{fig:sketch+phase-diagram}(b), it appears almost flat and hence only weakly dependent on \(\lambda_2\), its full behaviour is more intricate. Specifically, the boundary is a non-monotonic function of \(\lambda_2\). In the asymptotic regime \(\lambda_2 \gg \lambda_1\), it is approximately described by
$\lambda_1 \sim \left(\frac{\lambda_2}{16}\right)^{1/3}$ as shown in Fig.~\ref{fig:long_boundary}.

\begin{figure}[!h]
\centering
\includegraphics[scale = 0.45]{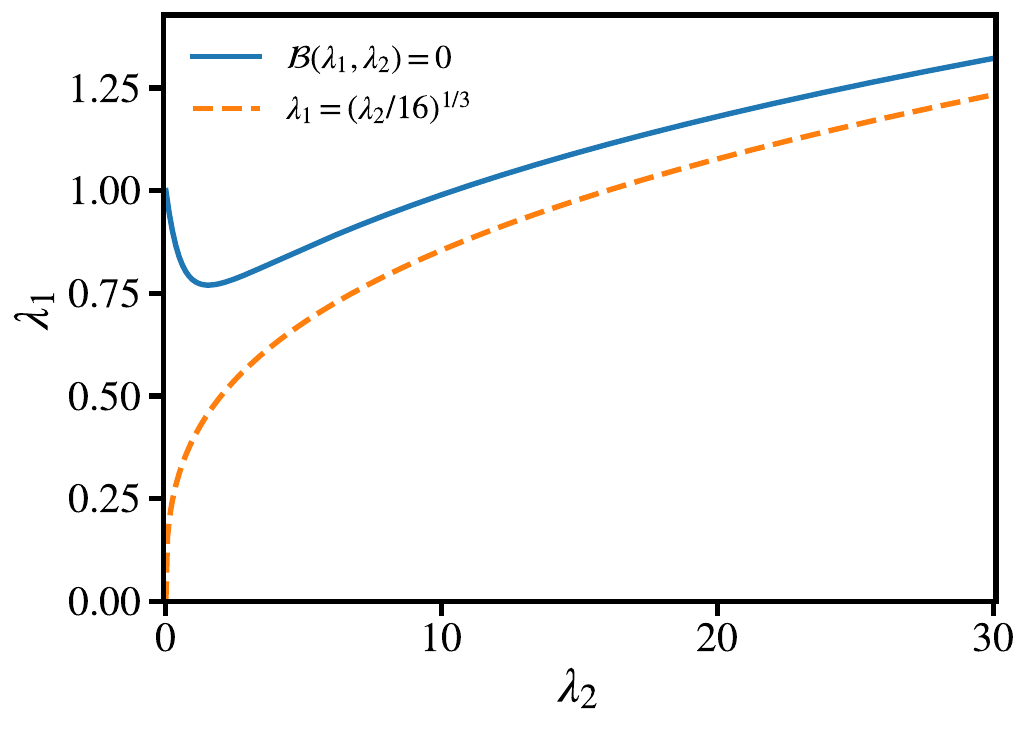}
\caption{Non-monotonous behaviour of boundary $\mathcal{B}(\lambda_1,\lambda_2)$.}
\label{fig:long_boundary}
\end{figure}
\FloatBarrier
\section{Stochastic quantum dynamics}
\label{sec:stoch-appendix}
\begin{figure*}[!t]
    \centering
    \includegraphics[scale=1]{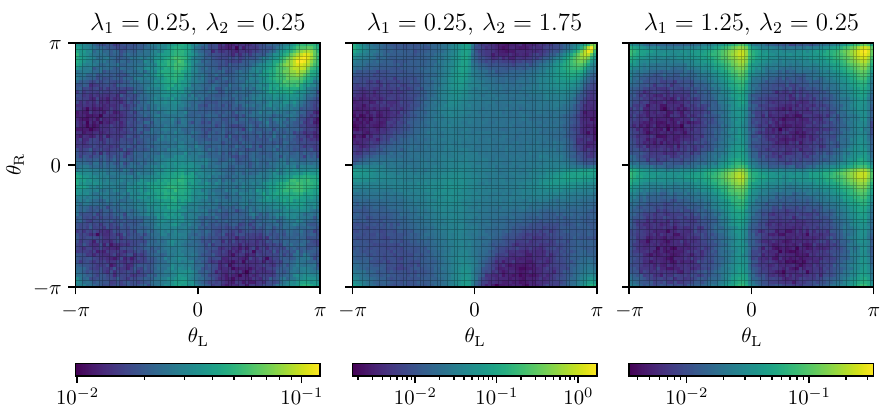}
    \caption{Stationary probability distribution \( P_\infty(\theta_L, \theta_R) \) for full simulation (evolution by SSE). The data for the exact simulation is obtained by simulating \( 10^6 \) trajectories that are stopped at \( T = 20 \) and whose starting point is the state \( |11\rangle \). The couple of angles at the final time are binned on a grid of size \( 72 \times 72 \).}
    \label{fig:stationary-probability-full_SSE}
\end{figure*}

\begin{figure*}[!t]
    \centering
    \includegraphics[scale=01]{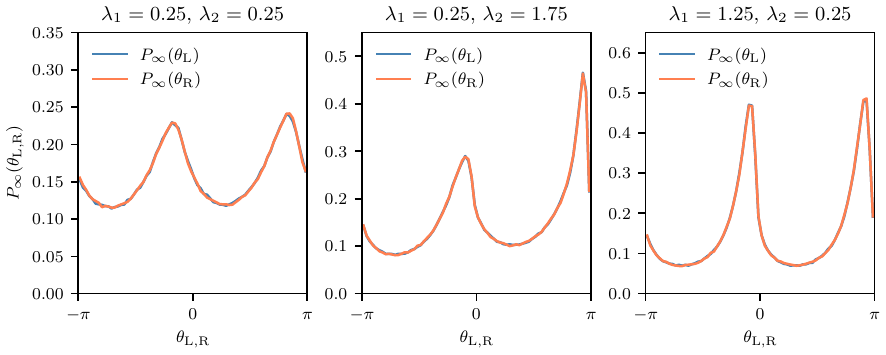}
    \caption{Stationary marginal probability distribution \( P_\infty(\theta_{L,R}) \) for full simulation (evolution SSE).}
    \label{fig:stationary-marginal-probability_SSE}
\end{figure*}

We present here the results for the PDF studied in the main text for a different measurement model, i.e. quantum state diffusion.
The model describes a continuous stochastic evolution of the quantum state, similar to that of the quantum jumps model in the main text, but with a continuous readout output (and associated Kraus operators) per measurement.
In this case, the evolution of the system is also described by quantum trajectories \( \ket{\psi(t)} \), which follow a Stochastic Schrödinger Equation (SSE) given by:
\begin{widetext}
\begin{equation}
\label{SSE}
d\ket{\psi(t)} =
-iH\,dt\,\ket{\psi(t)}
-\frac{dt}{2}\sum_{r=1}^{3}\gamma^{(r)}
\bigl(O^{(r)}-\langle O^{(r)}\rangle\bigr)^2 \ket{\psi(t)}
+\sum_{r=1}^{3} dW^{(r)}
\bigl(O^{(r)}-\langle O^{(r)}\rangle\bigr)\ket{\psi(t)} .
\end{equation}
\end{widetext}

where \(dW^{(r)}\) are uncorrelated stochastic increments sampled from a Gaussian distribution satisfying \(\langle dW^{r} \rangle = 0\) and \(\langle dW^{r} dW^{r'} \rangle = \gamma^{(r)} dt \ \delta_{r,r'} \). Here, \(\gamma^{(1)} = \gamma^{(2)} = \gamma_1\), \(\gamma^{(3)} = \gamma_2\) and the operators are defined as \(O^{(1)} = n_L\), \(O^{(2)} = n_R\), and \(O^{(3)} = n_L n_R\).

This framework differs from quantum jumps, which describe dynamics as a sequence of discrete, abrupt changes (jumps) interspersed with smooth deterministic evolution under a non-Hermitian Hamiltonian. In contrast, stochastic quantum dynamics focuses on gradual, continuous evolution influenced by noise, offering a different perspective on open quantum systems under continuous monitoring.

We numerically simulate the dynamics in \ref{SSE} and compute the stationary probability distribution $P_\infty(\theta_L,\theta_R)$. The results, shown in Fig.~\ref{fig:stationary-probability-full_SSE}, reveal that for weak measurements, the system under stochastic quantum dynamics behaves similarly to quantum jumps, whereas for strong measurements, the topology undergoes a qualitative change, and doesn't show any forbidden  (see Fig.\ref{fig:stationary-marginal-probability_SSE}).

\bibliography{bibliography}
\end{document}